\documentclass[conference]{IEEEtran}
\IEEEoverridecommandlockouts
\usepackage{cite}
\usepackage{amsmath,amssymb,amsfonts}
\usepackage{algorithmic}
\usepackage{graphicx}
\usepackage{textcomp}
\usepackage{xcolor}
\usepackage{url}
\usepackage[inline]{enumitem}
\usepackage{subcaption}
\usepackage{float}

\title{ A Resource-Efficient CNN-Based EEG Auditory Attention Decoding ASIC}

\author{
\IEEEauthorblockN{
    Qier Ma\IEEEauthorrefmark{1},
    Richard George\IEEEauthorrefmark{1},
    Stefan Scholze\IEEEauthorrefmark{1},
    Jehn Constantin\IEEEauthorrefmark{2},
    Tobias Reichenbach\IEEEauthorrefmark{2},
    and Christian Mayr\IEEEauthorrefmark{1}
}
\IEEEauthorblockA{
\IEEEauthorrefmark{1}Department of Electrical Engineering and Information Technology,\\
Institute of Circuits and Systems, Chair of Highly-Parallel VLSI Systems and Neuro-Microelectronics,\\
Technische Universit\"at Dresden, Dresden, Germany\\
Email: qier.ma@tu-dresden.de\\
\IEEEauthorrefmark{2}Department of Artificial Intelligence in Biomedical Engineering (AIBE),\\
Friedrich-Alexander-Universit\"at Erlangen–Nürnberg (FAU), Erlangen, Germany
}
\thanks{\textcopyright~2026 IEEE. Personal use of this material is permitted.
Permission from IEEE must be obtained for all other uses.}
}

\begin{document}
\maketitle

\begin{abstract}
Following a target speaker in a noisy environment, commonly known as the cocktail party problem, remains particularly challenging for cochlear implant (CI) users. Recent studies have explored EEG-based auditory attention decoding (AAD) using neural networks to enhance hearing assistance. This paper presents a resource-efficient ASIC for real-time EEG-based auditory attention decoding by integrating a quantized CNN inference engine and a Pearson-correlation classifier. The proposed architecture employs streaming execution, on-chip buffering, and memory-efficient dataflow to reduce hardware cost while maintaining real-time performance.

The proposed ASIC has been fully implemented in GF22FDX 22-nm CMOS technology, occupying a total silicon area of  2.09 mm$^2$(1264µm x 1654µm), with the CNN inference engine and streaming classification engine requiring only 0.076 mm$^2$. Operating at a core voltage of 0.55 V, the design achieves a power consumption of 0.4941 mW and an inference latency of 7.34 ms, providing an energy-efficient hardware platform for EEG-based auditory attention decoding in hearing-assistance applications.


\end{abstract}


\section{Introduction}

The cocktail party problem \cite{cocktail} remains a fundamental challenge for cochlear implant (CI) users in multi-speaker environments. Recent studies \cite{neural} have shown that electroencephalography (EEG) signals contain reliable neural correlates of auditory attention, enabling Auditory Attention Decoding (AAD) to identify the attended speaker directly from brain activity. The decoded attention information can provide a control signal for a hearing-assistance system to select or enhance the attended speech stream while suppressing competing speakers. 

Existing AAD approaches can be broadly divided into linear and neural-network-based methods. Linear techniques \cite{AAD_daily, lin, closedloop}, such as Canonical Correlation Analysis (CCA), are attractive for hardware-constrained real-time systems because of their low computational complexity and deterministic latency. In contrast, deep neural network (DNN)-based approaches \cite{AAD_env, AAD_dense} generally achieve higher decoding accuracy, but their computational complexity and memory requirements make efficient hardware implementation significantly more challenging.

Recent studies have also explored discriminative AAD approaches that directly classify auditory attention from EEG signals without stimulus reconstruction, enabling faster response times \cite{SGCN, ASAD_LSTM, ASAD1, ASAD2, ASAD3}. Representative methods include Auditory Spatial Attention Decoding (ASAD) \cite{TAnet, XAnet}, which exploits spatially asymmetric neural responses, as well as approaches incorporating visual information \cite{visual}, conversational dynamics \cite{turn}, or multi-speaker learning strategies \cite{4talker}. However, many of these approaches rely on spatial cues or predefined speaker configurations, which may limit their robustness in practical acoustic environments.

In contrast, stimulus reconstruction-based AAD remains independent of speaker locations and preserves the reconstructed speech envelope, which can serve as an auxiliary cue for speech separation frameworks \cite{BISS}. This property makes reconstruction-based AAD particularly attractive for practical hearing-assistance systems, especially in dynamic acoustic environments where speaker positions are unknown or continuously changing.

However, existing AAD studies primarily focus on decoding performance rather than efficient hardware implementation. This motivates compact architectures that satisfy the stringent power, area, and real-time constraints of practical hearing-assistance systems.

Motivated by these challenges, this work adopts the compact CNN architecture proposed in \cite{mike, constantin} and presents a resource-efficient ASIC integrating a quantized CNN inference engine with a Pearson-correlation classifier for real-time EEG-based auditory attention decoding. The proposed architecture employs streaming execution and memory-efficient dataflow to reduce hardware cost, and is implemented in GF22FDX 22-nm CMOS with post-layout verification.

\section{Hardware-Aware Algorithm Optimization}

\begin{figure}[thpb]  
    \centering
    \includegraphics[width=1.0\linewidth]{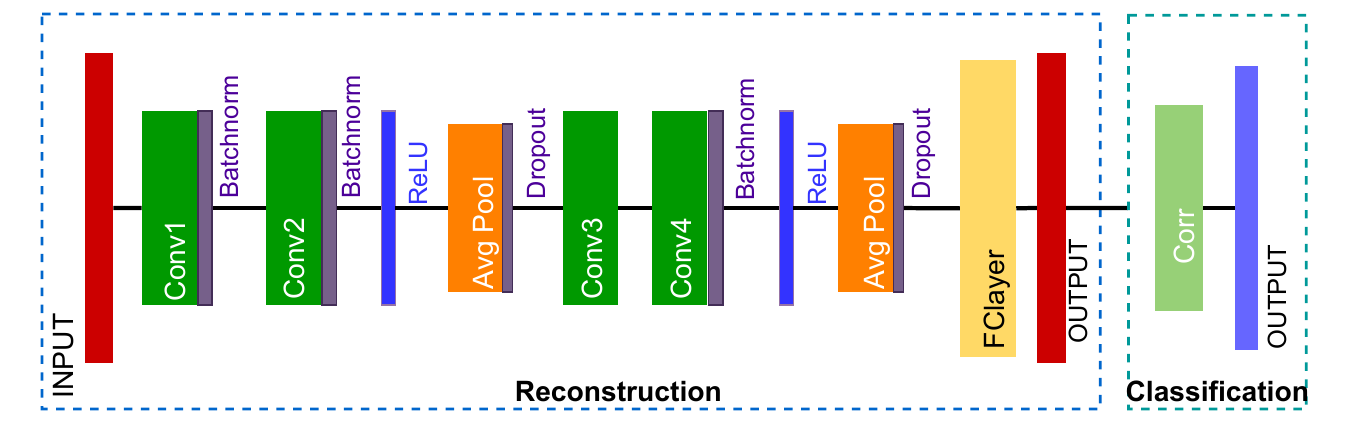}
    \caption{Proposed 4-layer Compact CNN Architecture}
    \label{fig:cnn}
\end{figure}

The chip performs EEG-based speech reconstruction to recover the attended audio envelope, followed by Pearson-correlation-based attention classification. The overall processing flow is illustrated in Fig.~\ref{fig:cnn}. Model training is performed offline, whereas the proposed ASIC executes real-time inference. Detailed network architecture and training procedures are described in \cite{mike,constantin}. To support subject-specific adaptation, model parameters can be updated through a serial SRAM interface between listening sessions.

The CNN processes a $31\times100$ EEG window. Rather than conventional layer-by-layer execution, the proposed design jointly optimizes algorithm transformation and hardware architecture to reduce memory traffic and hardware overhead.

To achieve this objective, several hardware-aware optimizations are applied. First, grouped convolution substantially reduces the number of trainable parameters and redundant MAC operations. Second, Batch Normalization (BN) folding absorbs the normalization parameters into the preceding convolution layers during offline compilation, eliminating dedicated normalization hardware during inference. Third, ReLU activation is adopted owing to its simple comparator-based implementation. Finally, an INT8 post-training quantization (PTQ) flow is employed, while Power-of-Two (PoT) scaling factors replace costly multiplication operations with bit shifts, further reducing arithmetic complexity and critical-path delay.

\section{Hardware Implementation}

\begin{figure}[htpb]  
    \centering
    \includegraphics[width=1.0\linewidth]{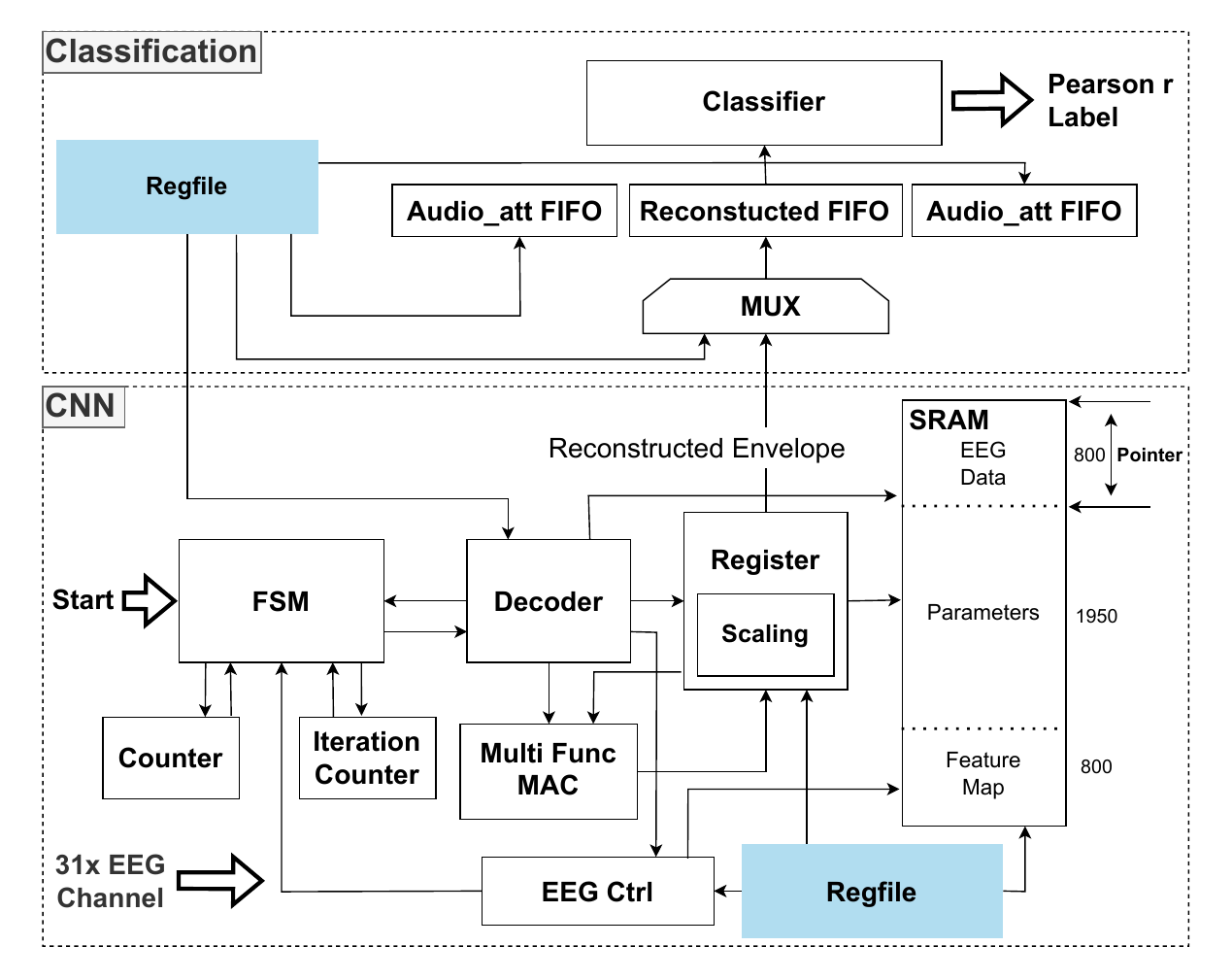}
    \caption{Overall architecture of the proposed EEG-AAD ASIC.}
    \label{fig:block}
\end{figure}

As illustrated in Fig.~\ref{fig:block}, the system generates a reconstructed stimulus from 31 EEG channels, once triggered by a start signal. Then, two correlation coefficients and a final attention label are computed as the output.
A cross-layer streaming execution flow is proposed to eliminate unnecessary feature-map materialization between adjacent CNN layers. By leveraging optimized loop ordering and lightweight buffering, intermediate feature maps are propagated directly through the pipeline, bypassing frequent SRAM accesses. In this design, only the second convolutional layer feature map requires SRAM storage, substantially reducing register and memory utilization while sustaining high processing bandwidth.

\subsection{CNN Inference Engine}



Unlike Application-Specific Instruction-set Processors (ASIPs) \cite{ASIP}, the proposed CNN engine specializes its streaming dataflow and execution schedule for the fixed EEG-AAD network, reducing control overhead while maximizing data reuse and enabling deterministic continuous operation.

The engine comprises an FSM for loop scheduling and pipeline control, a decoder for control-signal generation, and a register-based execution unit that performs arithmetic, stores intermediate results, and interfaces with the MAC units and on-chip SRAM.
\subsubsection{Real-Time EEG Stream Management}

Continuous EEG acquisition forms the foundation of the proposed streaming processor. A modulo-addressed circular buffer, enabling uninterrupted streaming without the memory duplication required by conventional ping-pong buffering. As illustrated in Fig.~\ref{fig:block}, the \textit{EEG CTRL} module manages a pointer register to update eight SRAM words per cycle, ensuring temporal ordering and real-time operation under strict latency constraints.

The on-chip SRAM is logically partitioned into three functional regions: 
\begin{enumerate*}[label=(\arabic*)]
    \item a real-time buffer for incoming 31-channel EEG samples; 
    \item a parameter memory for reconfigurable weights and biases to accommodate subject-specific fine-tuning; and 
    \item an intermediate storage area for feature maps. 
\end{enumerate*}

This organization enables continuous inference while minimizing memory footprint and SRAM access overhead.




\subsubsection{Cross-Layer Streaming Dataflow}\label{sec:pip}
\begin{figure}[htpb]  
    \centering
    \includegraphics[width=1.0\linewidth]{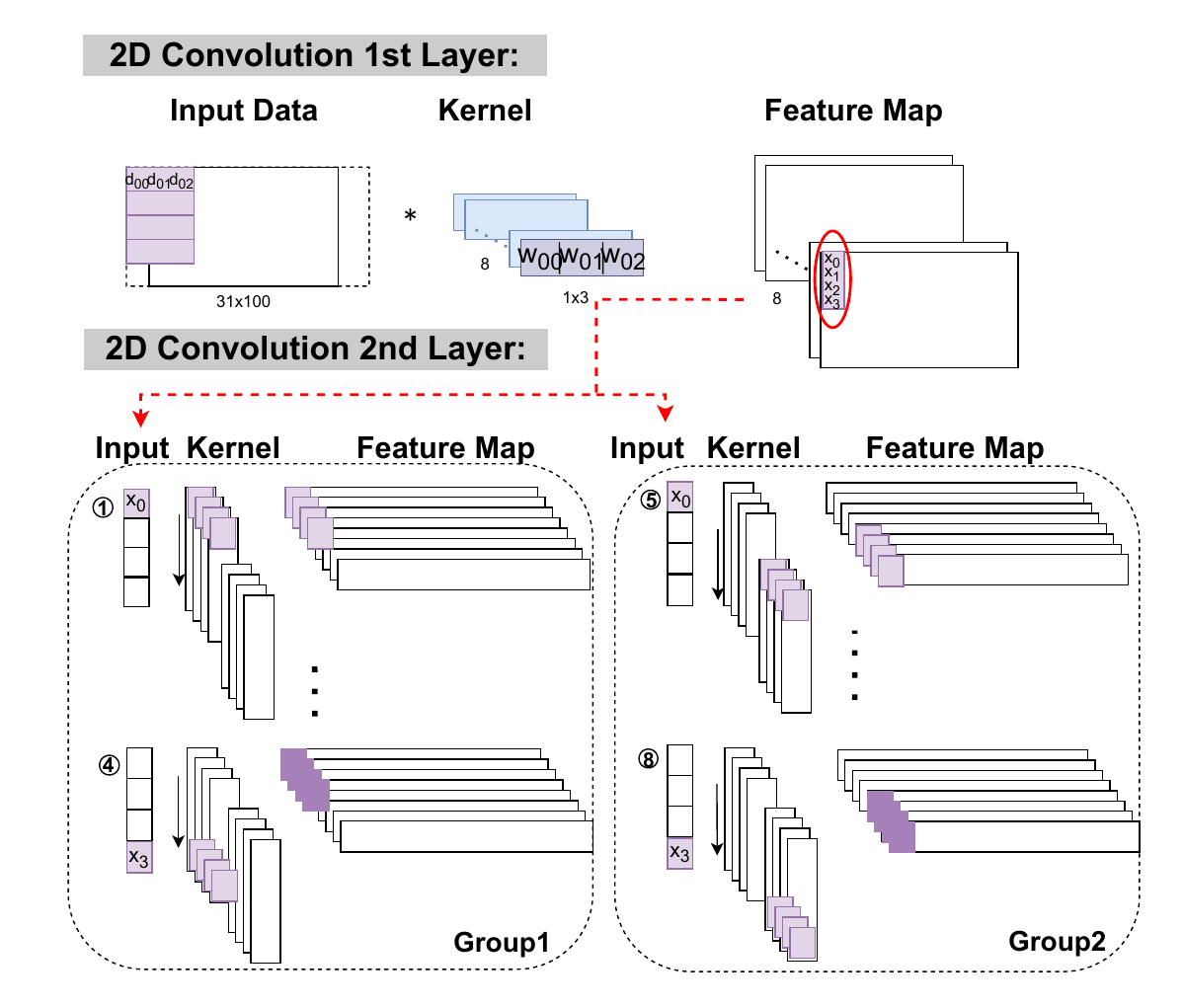}
    \caption{Parallel pipeline Example}
    \label{fig:pip}
\end{figure}

As illustrated in Fig.~\ref{fig:pip}, the computation engine is optimized for high hardware utilization and data reuse. To align the first convolution layer (kernel size [8,1,3]) with the SRAM organization, each Multi-MAC unit integrates four parallel multipliers to match the SRAM organization and maximize local data reuse. This configuration enables the concurrent execution of four channel-parallel operations, generating intermediate features across three accumulation cycles. These intermediate features are streamed directly into the second convolution layer, which is implemented as a grouped convolution (64 kernels). By adopting a synchronized execution order between layers, the design facilitates immediate data consumption and local accumulation.

Following accumulation, the feature maps undergo ReLU activation and average pooling (kernel size [2,1]). Through shift-based reduction, the final eight feature values are committed to on-chip SRAM. 
Consequently, only four registers for the first layer and eight for the second are required, completely eliminating full feature-map buffering while maintaining high throughput.






\subsection{Classification Module}

To preserve end-to-end streaming execution, the Pearson correlation is algebraically reformulated as:
\begin{equation}
\label{eq:pearson_sample}
r = \frac{\sum_{i=1}^n (x_i - \bar x)(y_i - \bar y)}
{\sqrt{\sum_{i=1}^n (x_i - \bar x)^2};\sqrt{\sum_{i=1}^n (y_i - \bar y)^2}}.
\end{equation}
This original formulation requires storing all samples until the mean values are computed, resulting in substantial memory overhead. To enable streaming computation, a rearranged sample-based form is adopted:
\begin{equation}
\label{eq:pearson_rearranged}
r = \frac{ n\sum_{i=1}^{n} x_i y_i - \sum_{i=1}^{n} x_i \sum_{i=1}^{n} y_i }
{\sqrt{ n\sum_{i=1}^{n} x_i^2 - \left(\sum_{i=1}^{n} x_i \right)^2 }
;\sqrt{ n\sum_{i=1}^{n} y_i^2 - \left(\sum_{i=1}^{n} y_i \right)^2 }}.
\end{equation}

This reformulation enables on-the-fly accumulation, eliminating the need to buffer complete signal histories.


To minimize memory overhead, only eight correlation coefficients are maintained in a circular buffer and updated once per second using a moving-average scheme.

The classifier addresses computational and synchronization challenges through two key strategies. First, area-efficient multi-cycle sequential operators are utilized for division and square-root functions, significantly reducing power and critical path compared to parallel architectures. Second, a FIFO-based synchronization mechanism ensures precise temporal alignment between the CNN-reconstructed signal and dual audio envelopes, triggering the classifier only when all input buffers are valid.

To reduce area and critical-path delay, division and square-root operations are implemented using multi-cycle sequential units. Since EEG inference operates at a low sampling rate, the resulting latency is naturally hidden within the streaming execution pipeline. A lightweight FIFO synchronizer aligns the reconstructed envelope with the two external audio envelopes, ensuring deterministic classifier activation only when all input streams are valid.



\section{Experimental Results and Discussion}
\subsection{Algorithmic Validation and Quantization}
\begin{figure}[htbp]
    \centering
    
    \begin{subfigure}[b]{1.0\linewidth}
        \centering
        \includegraphics[width=\linewidth]{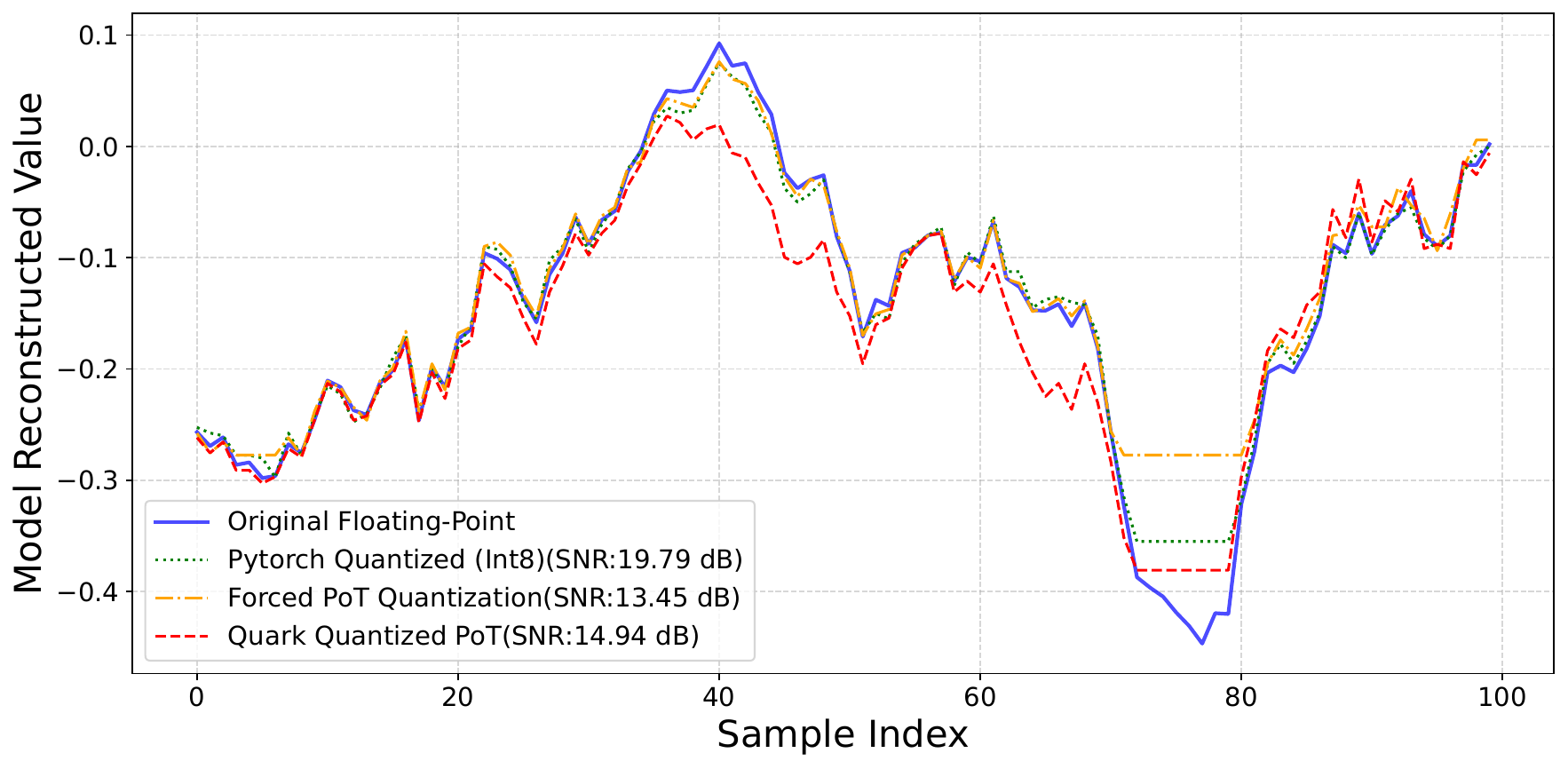}
        \caption{CNN Output (Audio Envelope) Reconstructed from EEG Signals}
        \label{fig:cnn_output}
    \end{subfigure}
    
    \vspace{0.5cm} 
    
    \begin{subfigure}[b]{1.0\linewidth}
        \centering
        \includegraphics[width=\linewidth]{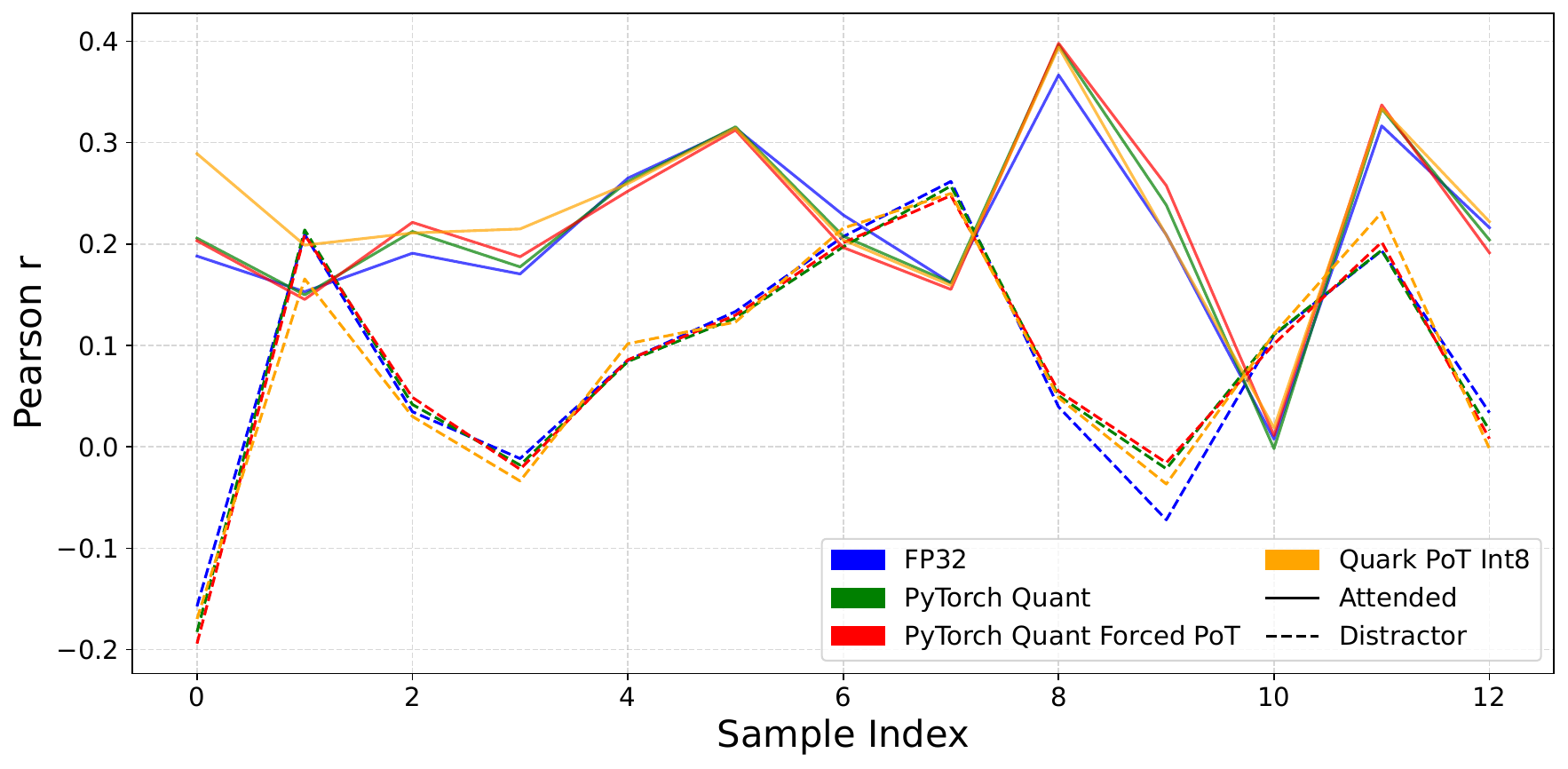}
        \caption{Correlation Coefficient}
        \label{fig:corr_plot}
    \end{subfigure}

    \caption{Hardware-oriented quantization evaluation: (a) provides the CNN stage performance, and (b) illustrates the final decoding accuracy.}
    \label{fig:combined_quantization}
\end{figure}

To evaluate the hardware-oriented optimization, the proposed architecture was validated using real EEG datasets collected by CI patients. The Floating-point 32 (FP32) model serves as the baseline for comparison. While this work primarily focuses on the physical hardware implementation (datasets, detailed algorithmic accuracy and architectural exploration are reported in our collaborative work \cite{constantin}), we emphasize the impact of Int8 Power-of-Two (PoT) quantization on hardware efficiency. The correlation coefficients (and, therefore, the final AAD accuracy) were remarkably robust to PoT quantization. This reflects the sensitivity of the Pearson correlation to an overall trend rather than individual datapoints.

To further explore the potential of PoT optimization, we investigated the AMD Quark PoT post-training quantization methodology. Experimental results reported in Fig.\ref{fig:combined_quantization} show that the Quark-optimized PoT model achieves an SNR improvement of approximately 1.5 dB compared to a baseline forced-replacement PoT model. As observed in this dataset, the attended correlation coefficients across all models predominantly exceed those of the distractor. 
We ultimately adopted the PyTorch-based quantization flow for the final hardware implementation due to its mainstream support and robust integration. 

Crucially, the hardware execution results are bit-true identical to the software-based PyTorch PoT simulation. This ensures seamless consistency between the algorithmic training phase and the final RTL implementation, 
guaranteeing that the hardware delivers the exact precision predicted during the software validation stage.

\subsection{Physical Implementation and Area Analysis}
The proposed neural processor has reached the tape-out stage and is implemented in  GF22FDX 22-nm CMOS technology. 

\begin{figure}[htbp]  
    \centering
    \includegraphics[width=1.0\linewidth]{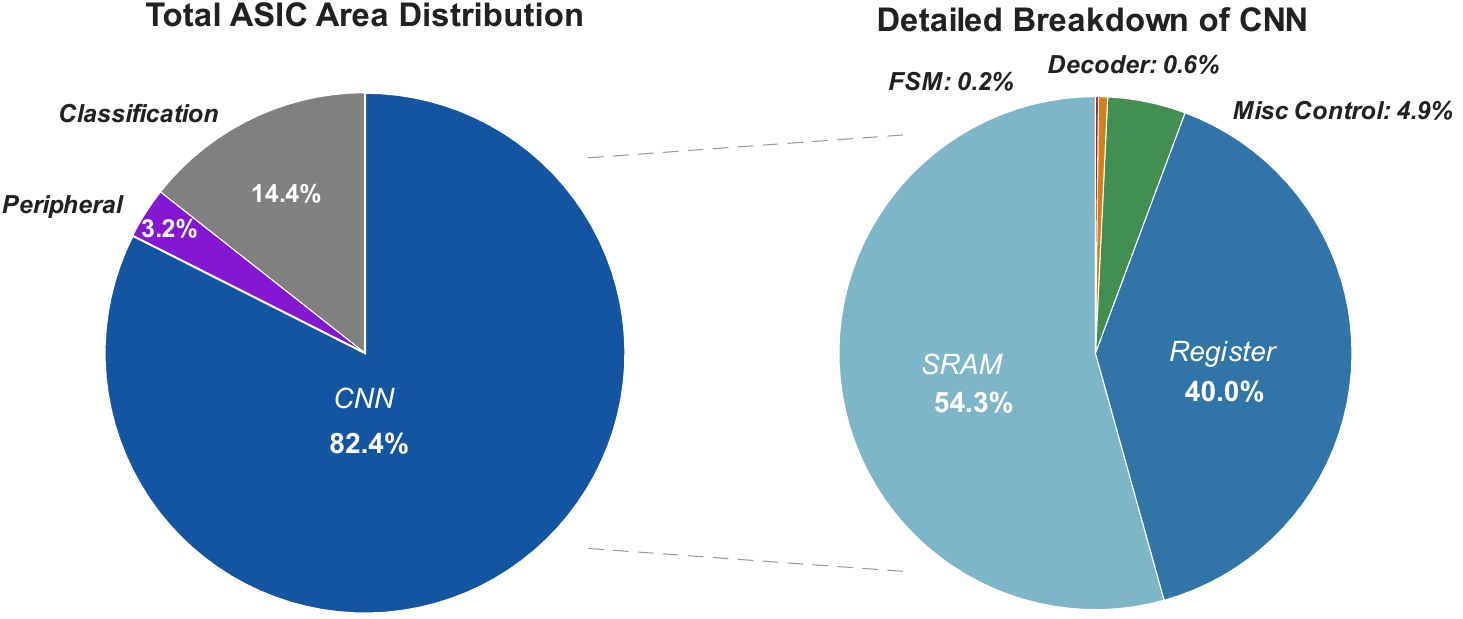}
    \caption{Area breakdown of the proposed streaming EEG-AAD processor.}
    \label{fig:area}
\end{figure}

The total silicon area is 76048 $\mu m^2$ with an instance count of 59149. As shown in Fig. \ref{fig:area}, the design is partitioned into three major components: the CNN Core, the Classifier, and the Peripherals (including UART and processor-to-core interfaces). 

The area distribution reflects the design philosophy of the proposed streaming processor. By eliminating repeated feature-map storage and employing highly serialized arithmetic, the computational logic occupies only a small fraction of the total area, while memory dominates the implementation.
\subsection{Power Evaluation and Energy Efficiency}

For post-layout power sign-off, we employed Cadence Voltus with gate-level VCD-based simulations. The results demonstrate a chip average power consumption of 0.4941 mW at the typical operating condition ($0.55\,\text{V}$, $25^\circ\text{C}$). 
Based on the hierarchical power analysis, the CNN and classification module account for 0.3265 mW.
To ensure thermal and functional reliability, the leakage power was further evaluated using Cadence Innovus under the specified power worst-case conditions ($0.6\,\text{V}$, $55^\circ\text{C}$), yielding an estimated leakage of 0.1953 mW.

The energy efficiency of the processor for a single inference is calculated as follows:
\begin{equation}
E_{\text{inf}} = P_{\text{total}} \times T_{\text{latency}} = 0.49\text{ mW} \times 7.34\text{ ms} =3.63\text{ }\mu\text{J/inf}
\end{equation}
The low power consumption is primarily achieved through the proposed streaming dataflow, which minimizes SRAM accesses and enables highly serialized arithmetic.
\subsection{Architecture Discussion and Trade-offs}

\begin{table}[htbp]
\centering
\caption{Comparison with Representative Hardware Implementations for EEG Processing}
\label{tab:comparison}
\resizebox{\columnwidth}{!}{
\begin{tabular}{|l|c|c|c|c|}
\hline
\textbf{Metric} & \textbf{EEGNet\cite{EEGnet}} & \textbf{HDC\cite{HDC}} & \textbf{SaleNet \cite{Salenet}} & \textbf{This Work} \\ \hline
\textbf{Method} & CNN & HDC & CNN & \textbf{CNN+Classifier} \\ \hline
\textbf{Platform} & FPGA (65nm) & FPGA (16nm) & FPGA (Artix-7) & \textbf{22nm ASIC} \\ \hline
\textbf{Application} & General EEG & Emotion Recog. & Attention Level & \textbf{AAD} \\ \hline
\textbf{Parameters} & $\sim$3,000 & 4 kbits & 30.91 k & \textbf{$\sim$7,000} \\ \hline
\textbf{Delay} & 24.4 ms & 628 ns & 2.01 ms & \textbf{7.34 ms} \\ \hline
\textbf{Energy (mJ/Inf)} & 0.267 & $4.2 \times 10^{-5}$ & $\sim$0.22 & \textbf{$3.63 \times 10^{-3}$} \\ \hline
\textbf{Memory (kB)} & 49.88 & $\sim$18 & $>$128 (1 Mb) & \textbf{11 } \\ \hline
\textbf{Area ($\text{mm}^2$)} & N/A & 0.596 & N/A & \textbf{0.076048} \\ \hline
\end{tabular}
}
\end{table}

As summarized in Table~\ref{tab:comparison}, direct comparison should be interpreted with caution due to differences in applications, workloads, and system functionality. For example, HDC-based processors \cite{HDC} achieve extremely high energy efficiency through binary hyperdimensional computing, whereas CNN-based processors provide substantially stronger feature extraction capability required for auditory attention decoding. Therefore, the comparison mainly highlights hardware efficiency across representative EEG processing architectures rather than absolute algorithmic superiority.

Among CNN-based implementations \cite{EEGnet,Salenet}, the proposed architecture demonstrates significant advantages in power consumption. Although the inference latency is slightly higher than that of SaleNet \cite{Salenet}, it remains well within the real-time requirement of a 125 Hz EEG stream (8 ms/sample). Rather than maximizing computational parallelism, the processor matches hardware resources to the EEG sampling rate through streaming execution and serialized arithmetic, achieving real-time throughput with substantially reduced hardware overhead.


\section{Conclusion}
This paper presents a resource-efficient streaming EEG-AAD processor integrating a quantized CNN inference engine and a Pearson-correlation classifier.
 The proposed architecture, implemented in GF22FDX 22-nm CMOS technology, achieves a processing latency of 7.34 ms. 
The average power consumption is 0.4941 mW under the typical corner ($0.55\text{V}$, $25^\circ\text{C}$).
The design demonstrates superior suitability for energy-constrained biomedical applications. 

The chip has been successfully taped out. Post-layout simulation results demonstrate robust performance across different PVT corners. Our future work will focus on the silicon characterization of the fabricated prototypes, including measured power analysis and real-time validation with patient-derived EEG data to further evaluate the system-level performance.

\section{Acknowledgment}
This project was supported by the German Federal Ministry of
Research, Technology and Space (Cluster4Future, SEMECO, project number 03ZU1210FA). We also express our sincere gratitude to Johannes Partzsch for the guidance and support throughout the manuscript preparation and writing process.

\newpage
\bibliographystyle{IEEEtran}
\bibliography{reference}

\end{document}